\documentclass[12pt,twoside]{scrartcl}
\usepackage{geometry}
\usepackage[latin1]{inputenc}
\usepackage[english]{babel}
\usepackage{amsmath,amssymb}
\usepackage{theorem}
\usepackage{a4}
\usepackage{epsfig}
\usepackage[hyphens]{url}
\usepackage[gen]{eurosym}

\begin{document}

\author{
        \large Maximilian Speicher \\
        \large Technische Universit\"at Chemnitz \\
        \large Fakult\"at f\"ur Informatik \\
        \large maximilian.speicher@s2013.tu-chemnitz.de}
\title{What is Usability?}
\subtitle{A Characterization based on ISO 9241-11 and ISO/IEC 25010}
\date{}

\maketitle

\begin{abstract}
\textbf{Abstract.} According to Brooke~\cite{Brooke96} ``Usability does not exist in any absolute sense; it can only be defined with reference to particular contexts.''
That is, one cannot speak of usability without specifying what \emph{that} particular usability is characterized by.
Driven by the feedback of a reviewer at an international conference, I explore in which way one can precisely specify the kind of usability they are investigating in a given setting.
Finally, I come up with a formalism that defines usability as a quintuple comprising the elements \textit{level of usability metrics, product, users, goals} and \textit{context of use.}
Providing concrete values for these elements then constitutes the investigated type of usability.
The use of this formalism is demonstrated in two case studies.
\end{abstract}

\section{Introduction}
\label{sec:introduction}

In 2014, I submitted a research paper about a concept called \textit{Usability-based Split Testing}\footnote{``Usability-based Split Testing'' means comparing two variations of the same web interface based on a quantitative usability score (e.g., usability of interface A = 97\%, usability of interface B = 42\%)~\cite{Speicher-ICWE14}. The split test can be carried out as a user study or under real-world conditions~\cite{Speicher-ICWE14}.} to a web engineering conference~\cite{Speicher-ICWE14}.
My evaluation involved a questionnaire that asked for ratings of different factors of usability based on a novel usability instrument specifically developed for web interfaces~\cite{Speicher-DUXU15}.
This instrument comprises the items \textit{informativeness, understandability, confusion, distraction, readability, information density} and \textit{reachability}, which have been identified as factors of usability in a confirmatory factor analysis~\cite{Speicher-DUXU15}.
So obviously, I use the word ``usability'' in that paper a lot; however, without having thought of its exact connotation in the context of my research before.
Of course I was aware of the differences compared to User eXperience (UX; cf.\ \cite{Law-CHI09}), but just assumed that the used questionnaire and description of my analyses would make clear what my paper understands as usability.

Then came the reviews and one reviewer noted:

\begin{quote}
``There is a weak characterization of what Usability is in the context of Web Interface Quality, quality models and views.
Usability in this paper is a key word.
However, it is weakly defined and modeled w.r.t.\ quality.''
\end{quote}

This confused me at first since I thought it was pretty clear what usability is and that my paper was pretty well understandable in this respect.
In particular, I thought \textit{Usability has already been defined and characterized before, so why does this reviewer demand me to characterize it again?}
Figuratively speaking, they asked me: ``When you talk about usability, what is that \frq usability\flq?''

\section{A Definition of Usability}

As I could not just ignore the review, I did some more research on definitions of usability.
I remembered that Nielsen defined usability to comprise five quality components---Learnability, Efficiency, Memorability, Errors, and Satisfaction~\cite{Nielsen12}.
Moreover, I had already made use of the definition given in ISO 9241-11~\cite{ISO9241-11} for developing the usability questionnaire (cf.\ \cite{Speicher-DUXU15}) used in my evaluation:

\begin{quote}
``The extent to which a product can be used by specified users to achieve specified goals with effectiveness, efficiency and satisfaction in a specified context of use.''~\cite{ISO9241-11}
\end{quote}

During the design of the questionnaire I had focused only on reflecting the mentioned high-level factors of usability---effectiveness, efficiency, and satisfaction---by the contained items.
However, the rest of the definition is not less interesting. Particularly, it contains the phrases

\begin{enumerate}
 \item ``a product'';
 \item ``specified users'';
 \item ``specified goals''; and
 \item ``specified context of use''.
\end{enumerate}

As can be seen, the word ``specified'' is used three times---and also ``a product'' is a rather vague description here.

This makes it clear that usability is a difficult-to-grasp concept and even the ISO definition~\cite{ISO9241-11} gives ample scope for different interpretations.
Also, in his paper on the \textit{System Usability Scale}, Brooke~\cite{Brooke96} refers to ISO 9241-11 and notes that ``Usability does not exist in any absolute sense; it can only be defined with reference to particular contexts.''
Thus, one has to explicitly \emph{specify} the four vague phrases mentioned above to characterize the exact manifestation of usability they are referring to.
Despite my initial skepticism, that reviewer was absolutely right.

While usability is of course also an attribute of everyday things such as doors or coffee machines, in this technical report the fundamental assumption is that we are talking about settings that involve interfaces provided by visual displays, which is based on and in accordance with ISO 9241-11~\cite{ISO9241-11}.

\section{Levels of Usability Metrics}

As the reviewer explicitly referred to ``Web Interface Quality'', we also have to take ISO/IEC 25010~\cite{ISO25010} (that has replaced ISO/IEC 9126) into account.
That standard is concerned with \textit{software engineering} and \textit{product quality} and, among other things, refers to three different levels of quality metrics~\cite{ISO25010}:

\begin{itemize}
 \item \textit{Internal metrics,} which measure a set of static attributes (e.g., related to software architecture and structure).
 \item \textit{External metrics,} which relate to the behavior of a system (i.e., they rely on execution of the software).
 \item \textit{In-use metrics,} which involve actual users in a given context of use.
\end{itemize}

ISO/IEC 25010 defines usability as a subset of \textit{quality in use}~\cite{ISO25010}, which makes sense as ``usability'' is derived from the word ``use'' and cannot be meaningfully applied to products that are not actually used.
Yet, it is possible to draw inferences about usability from static attributes and measures that rely on software execution alone.
Hence, we transfer the three types of metrics above into the context of usability evaluation.
In analogy, this gives us three levels of usability metrics: \textit{Internal usability metrics, external usability metrics,} and \textit{usability in use metrics}.

This means that if we want to evaluate usability, we first have to state which of the above levels we are investigating.
The first one (internal usability metrics) might be assessed with a static code analysis, as for example carried out by accessibility tools that among other things check whether the \texttt{alt} attributes of all images are set on a webpage.
The second (external usability metrics) might be assessed in terms of an expert going through a rendered interface without actually using the product, or as is done by \textit{jQMetrics}\footnote{\url{https://github.com/globis-ethz/jqmetrics}, retrieved January 22, 2015.}.
Finally, usability in use metrics are commonly assessed with user studies, either on a live website, or in a more controlled setting.

\section{Bringing it all Together}

Once we have decided for one of the above levels of usability metrics, we have to give further detail on the four vague phrases contained in ISO 9241-11~\cite{ISO9241-11}.
Mathematically speaking, we have to find concrete values for the elements \textit{product}, \textit{users}, \textit{goals}, and \textit{context of use}, which are sets of characteristics.
Together with the level of usability metrics, this gives us a quintuple defined by the following Cartesian product: 

$$
\textit{usability} \in \textit{LEVEL} \times \textit{PRODUCT} \times \textit{USERS} \times \textit{GOALS} \times \textit{CONTEXT}
$$

We already know the possible values for \textit{level of usability metrics}:

\begin{equation}
\begin{split}
\textit{level of usability metrics} \in \textit{LEVEL} \\
\textit{LEVEL} = \{\text{internal}, \text{external}, \text{in use}\}
\end{split}
\end{equation}

So what are the possible values for the remaining elements contained in the ``quintuple of usability''?

\subsection{Product}

The first one is rather straightforward.
\textit{Product} is the actual product you are evaluating, or at least the type thereof.
Particularly, web interface usability is different from desktop software or mobile app usability.
Also, it is important to state whether one evaluates only a part of an application (e.g., a single webpage contained in a larger web app), or the application as a whole.
Therefore: 

\begin{equation}
\begin{split}
\textit{product} \subseteq \textit{PRODUCT} \\
\textit{PRODUCT} = \{\text{desktop application}, \text{mobile application}, \text{web application}, \\
\text{online shop}, \text{WordPress blog}, \text{individual web page}, ...\}
\end{split}
\end{equation}

Since \textit{product} is a subset of the potential values, it is possible to use any number of them for a precise characterization of the element.
For instance, $\textit{product} = \{\text{mobile application}, \text{WordPress blog}\}$ if you are evaluating the mobile version of your blog.
This should not be thought of as a strict formalism, but is rather intended as a convenient way to express the combined attributes of the element.
However, not all values can be meaningfully combined (e.g., desktop application and WordPress blog).
Therefore, the correct definition and usage are the responsibility of the evaluator.\footnote{In this case, ``evaluator'' means the person who has to specify the considered type of usability. This can also include stakeholders, product owners, developers etc.}
The same holds for the remaining elements explained in the following.

\subsection{Users}

Next comes the element \textit{users}, which relates to the target group of your product (if evaluating in a real-world setting) or the participants involved in a controlled usability evaluation (such as a lab study).
To distinguish between these is highly important since different kinds of users perceive a product completely differently.
Also, real users (preferably in a real-world setting) are more likely unbiased compared to participants in a usability study.

\begin{equation}
\begin{split}
\textit{users} \subseteq \textit{USERS} \\
\textit{USERS} = \{\text{visually impaired users}, \text{female users}, \text{users aged 19--49}, \\
\text{test participants}, \text{inexperienced users}, \text{experienced users}, \text{novice users}, \\
\text{frequent users}, ...\}
\end{split}
\end{equation}

In particular, when evaluating usability in a study with participants, this element should contain all demographic characteristics of that group.
Yet, when using methods such as expert inspections (cf.\ \cite{Wixon94}), \textit{users} should not contain ``usability experts,'' as your interface is most probably not exclusively designed for that very specific group.
Rather, it contains the characteristics of the target group the expert has in mind when performing, for instance, a cognitive walkthrough (cf.\ \cite{Wharton94}).
This is due to the fact that usability experts are usually well-trained in simulating a user with specific attributes.

\subsection{Goals}

The next one is a bit tricky, as \textit{goals} are not simply the tasks a specified user shall accomplish (such as completing a checkout process).
Rather, there are two types of goals according to Hassenzahl~\cite{Hassenzahl-IHM08}: \textit{do-goals} and \textit{be-goals}. 

\textit{Do-goals} refer to the \textit{pragmatic} dimension, which means ``the product's perceived ability to support the achievement of [tasks]''~\cite{Hassenzahl-IHM08}, as for example the aforementioned completion of a checkout process.

Contrary, \textit{be-goals} refer to the \textit{hedonic} dimension, which ``calls for a focus on the Self''~\cite{Hassenzahl-IHM08}.
To give just one example, the ISO 9241-11~\cite{ISO9241-11} definition contains ``satisfaction'' as one component of usability.
Therefore, ``feeling satisfied'' is a be-goal that can be achieved by users.
The achievement of be-goals must not necessarily be connected to the achievement of corresponding do-goals, i.e.\ do-goals are not inevitably a prerequisite~\cite{Hassenzahl-IHM08}.
This means that a user can be satisfied even if they failed to accomplish certain tasks and vice versa~\cite{Hassenzahl-IHM08}.

Thus, it is necessary to take these differences into account when defining the specific goals to be achieved by a user.
The element \textit{goals} can be specified either by the concrete tasks the user shall achieve or by Hassenzahl's~\cite{Hassenzahl-IHM08} more general notions if no specific tasks are defined:

\begin{equation}
\begin{split}
\textit{goals} \subseteq \textit{GOALS} \\
\textit{GOALS} = \{\text{do-goals}, \text{be-goals}, \text{completed checkout process}, \\
\text{writing a blog post}, \text{feeling satisfied}, \text{having fun}, ...\}
\end{split}
\end{equation}

Particularly, the dimensions of usability given by ISO 9241-11~\cite{ISO9241-11}---effectiveness, efficiency and satisfaction---can be expressed by elements of the set \textit{GOALS}: ``being effective'', ``being efficient'' and ``being satisfied''.

For more information about \emph{goal-directed design}, the interested ready may refer to~\cite{Cooper04}.

\subsection{Context of use}

Last comes the element \textit{context of use}.
This one describes the setting in which you want to evaluate the usability of your product.
In particular, context is strongly connected to device-related differences, e.g.,\ a desktop PC vs.\ a touch device.
Recently, British newspaper \textit{The Guardian} reported their website is accessed by 6000 different types of devices per month.\footnote{\url{http://next.theguardian.com/blog/responsive-takeover/}, retrieved January 25, 2015.}
However, it is not sufficient to define context only by the device used.
It also contains more general information about the setting---such as ``real world'' or ``lab study'' to indicate a potential bias of the users involved---, user-related properties and other more specific information.
For instance, if you are evaluating the usability of a location-based service, your context most probably includes mobile devices that are used outside, i.e.\ with a potentially \textit{higher noise level} than at home, \textit{suboptimal light conditions} and a potentially \textit{weak signal strength}.
In~\cite{Dey-PUC01}, Dey defines context as follows:

\begin{quote}
``Context is any information that can be used to characterize the situation of an entity. An entity is a person, place, or object that is considered relevant to the interaction between a user and an application, including the user and applications themselves.''
\end{quote}

In general, your setting/context should be described as precisely as possible. 

\begin{equation}
\begin{split}
\textit{context of use} \subseteq \textit{CONTEXT} \\
\textit{CONTEXT} = \{\text{real world}, \text{lab study}, \text{expert inspection}, \text{desktop PC}, \\
\text{mobile phone}, \text{tablet PC}, \text{at day}, \text{at night}, \text{at home}, \text{at work}, \text{user is walking}, \\
\text{user is sitting}, ...\}
\end{split}
\end{equation}

\section{Case Studies}

\subsection{Evaluation of a Search Engine Results Page}

For testing a research prototype in the context of my industrial PhD thesis, we have evaluated a novel \textit{search engine results page} (SERP) designed for use with desktop PCs~\cite{Speicher-ICWE14}.
The test was carried out as a remote asynchronous user study with participants being recruited via internal mailing lists of the cooperating company.
They were asked to find a birthday present for a good friend that costs not more than \euro 50, which is a semi-open task (i.e., a do-goal).
According to our above formalization of usability, the precise type of usability $u$ assessed in that evaluation is therefore given by the following (for the sake of readability, the quintuple is given in list form):\footnote{As I have defined \textit{usability} in terms of a quintuple and tuples are ordered lists of elements, the formally correct notation would be: $u$ = $\big($usability in use, \{web application, SERP\}, \{company employees, novice users, experienced searchers (several times a day), average age $\approx$ 31, 62\% male, 38\% female\}, \{formulate search query, comprehend presented information, identify relevant piece(s) of information\}, \{desktop PC, HD screen, at work, remote asynchronous user study\}$\big)$.} 

\begin{itemize}
 \item \textit{level of usability metrics} = in use
 \item \textit{product} = \{web application, SERP\}
 \item \textit{users} = \{company employees, novice users, experienced searchers (several times a day), average age $\approx$ 31, 62\% male, 38\% female\}
 \item \textit{goals} = \{formulate search query, comprehend presented information, identify relevant piece(s) of information\}
 \item \textit{context of use} = \{desktop PC, HD screen, at work, remote asynchronous user study\}
\end{itemize}

In case the same SERP is inspected by a team of usability experts in terms of screenshots, the assessed type of usability changes accordingly.
In particular, \textit{users} changes to the actual target group of the web application, as defined by the cooperating company and explained to the experts beforehand.
Also, \textit{goals} must be reformulated to what the experts pay attention to (only certain aspects of a system can be assessed through screenshots).
Overall, the assessed type of usability is then expressed by the following:

\begin{itemize}
 \item \textit{level of usability metrics} = external
 \item \textit{product} = \{web application, SERP\}
 \item \textit{users} = \{German-speaking Internet users, any level of searching experience, age 14--69\}
 \item \textit{goals} = \{identify relevant piece(s) of information, be satisfied with presentation of results, feel pleased by visual aesthetics\}
 \item \textit{context of use} = \{desktop PC, screen width $\geq$ 1225 px, expert inspection\}
\end{itemize}

\subsection{A New Usability Instrument for Interface Evaluation}

In~\cite{Speicher-DUXU15} we describe the development of \textsc{Inuit}---a new usability instrument for interface evaluation.
As has already been mentioned in Section~\ref{sec:introduction}, \textsc{Inuit} comprises the seven items \textit{informativeness, understandability, confusion, distraction, readability, information density} and \textit{reachability}, which have been identified as factors of usability in a confirmatory factor analysis.
Yet, while such a limited set of items also has its advantages, it narrows the types of usability that can be investigated in settings based on this particular instrument.
Thus, the possible types of usability that can be evaluated are narrowed down as is explained in the following:

\begin{itemize}
 \item \textit{level of usability metrics:} The instrument is not suited for evaluations based on internal usability metrics, as items such as, e.g., \textit{readability} or \textit{distraction} can only be meaningfully judged with respect to the rendered interface. Thus, in this case \textit{level of usability metrics} $\in$ \{external, in use\}.
 \item \textit{product:} Using the instrument does not affect the types of products that can be evaluated, as long as they involve visual displays, which is a fundamental assumption in this technical report based on ISO 9241-11~\cite{ISO9241-11}. Therefore, \textit{product} $\subseteq$ \textit{PRODUCT}.
 \item \textit{users:} Using the instrument does not imply restrictions on the types of users an investigated interface targets. Therefore, \textit{users} $\subseteq$ \textit{USERS}.
 \item \textit{goals:} As the instrument assesses seven specific factors of usability, the investigated goals are limited and directly defined by the instrument's items, i.e., ``finding a desired piece of information'', ``understanding the presented information'', ``not being confused'' etc.\ and/or more fine-grained goals that are prerequisites for these (based on the specific interface that is investigated). Moreover, the dimension \textit{satisfaction}, which corresponds to goals such as ``feeling satisfied'', is not considered by the instrument in accordance with~\cite{Lew-ICWE10}. Based on an assumption like ``users are only satisfied when they found their desired piece of information'', one could still try to infer satisfaction from the given items. However, the instrument does not directly ask users whether they were satisfied. Therefore, \textit{goals} $\subseteq$ \{finding a desired piece of information, understanding the presented information, not being confused, not being distracted, ...\}.
 \item \textit{context of use:} Using the instrument does not affect the types of contexts that can be evaluated. Therefore, \textit{context of use} $\subseteq$ \textit{CONTEXT}.
\end{itemize}

\section{Conclusion}

Usability is a term that spans a wide variety of potential manifestations.
For example, usability evaluated in a real-world setting with real users might be a totally different kind of usability than usability evaluated in a controlled lab study---even with the same product.
Therefore, a given set of characteristics must be specified or otherwise, the notion of ``usability'' is rather meaningless due to its high degree of ambiguity.
It is necessary to provide specific information on five elements that have been identified based on ISO 9241-11~\cite{ISO9241-11} and ISO/IEC 25010~\cite{ISO25010}: \textit{level of usability metrics, product, users, goals,} and \textit{context of use.}
This has been demonstrated in two case studies based on existing research.
Although I have introduced a mathematically seeming formalism for characterizing the precise type of usability one is assessing, it is not necessary to provide that information in the form of a quintuple.
Rather, my primary objective is to raise awareness for careful specifications of usability, as many reports on usability evaluations---including the original version of my research paper~\cite{Speicher-ICWE14}---lack a complete description of what they understand as \frq usability\flq.

\subsection*{Acknowledgments}

Special thanks go to J\"urgen Cito, Sebastian Nuck, Sascha Nitsch, Tim Church \& my brother Frederic, who provided feedback on drafts of this technical report.


\bibliographystyle{plain}
\bibliography{refs}

\section*{Author Statement}

Earlier versions of this technical report have been originally published as ``What is \frq Usability\flq?'' on my personal blog \textit{2008} (or \textit{Twenty Oh Eight}): \url{http://2008.maxspeicher.com/2014/10/09/what-is-usability/}; and on \textit{Medium}: \url{https://medium.com/@maxspeicher/what-is-usability-bf578c2a772d}.

\end{document}